\documentclass[12pt]{article}
\usepackage{amsmath}
\usepackage{amssymb}
\usepackage[dvips]{epsfig}

\def\##1{{\bf #1}}
\def\=#1{\underline{\underline{ #1}}}

\def\*#1{\underline{\underline{\bf #1}}}

\def\skipline{\vskip 5mm}

\def\le{\left(}
\def\ri{\right)}
\def\les{\left[}
\def\ris{\right]}
\def\lec{\left\{}
\def\ric{\right\}}

\def\r#1{(\ref{#1})}

\def\.{\mbox{ \tiny{$^\bullet$} }}

\def\epso{\epsilon_{\scriptscriptstyle 0}}
\def\lambdao{\lambda_{\scriptscriptstyle 0}}
\def\muo{\mu_{\scriptscriptstyle 0}}

\def\eps{\epsilon}
\def\epsa{\epsilon_a(\omega)}
\def\epsb{\epsilon_b(\omega)}
\def\epsc{\epsilon_c(\omega)}
\def\epsg{\epsilon_g(\omega)}

\def\epsref{\=\epsilon_{\, ref}(\omega)}
\def\muref{\=\mu_{\, ref}(\omega)}
\def\alpharef{\=\alpha_{\, ref}(\omega)}
\def\betaref{\=\beta_{\, ref}(\omega)}

\def\utau{\#u_{\tau}}
\def\ux{\#u_x}
\def\uy{\#u_y}
\def\uz{\#u_z}
\def\ub{\#u_b}
\def\un{\#u_n}
\def\ur{\#u_r}

\def\Er{\#E(\#r,\omega)}
\def\Hr{\#H(\#r,\omega)}
\def\DDr{\#D(\#r,\omega)}
\def\Br{\#B(\#r,\omega)}

\begin{document}



\begin{center}
{\bf SCULPTURED THIN FILMS:\\ ACCOMPLISHMENTS
AND EMERGING USES}

\skipline
\skipline

Akhlesh Lakhtakia\footnote{Tel. +1 814 863 4319; Fax. +1 814 863 7967;
E--mail: AXL4@psu.edu}

\skipline
CATMAS --- Computational and Theoretical Materials Sciences Group,\\
Department of Engineering Science and Mechanics,\\
Pennsylvania State University, University Park, PA 16802--6812, USA

\end{center}
\skipline

 \vskip 2mm
{\bf Abstract}
\skipline
Sculptured thin films (STFs) are nano--engineered materials whose columnar
morphology is tailored to elicit desired optical responses upon excitation. Two
canonical forms of STFs have been identified. Linear constitutive relations for general
STFs as unidirectionally nonhomogeneous (continuously or piecewise uniformly)
and locally bianisotropic continuums are presented, along with a 4$\times$4 matrix
ordinary differential equation for wave propagation therein. A nominal model
for the macroscopic properties of linear STFs is devised from
nanoscopic considerations. The accomplished implementation
of STFs as circular polarization filters and spectral hole filters is discussed,
as also are emerging applications such as bioluminescence sensors
and optical interconnects.

\skipline
{\bf Keywords:} Bianisotropic materials,  Biochips, Chiral optics,
Local homogenization, Optical devices, Sculptured thin films, Sensors
\vskip 2.5mm

{\bf 1. Introduction}
\skipline

Shortly after the conceptualization of helicoidal bianisotropic mediums (HBMs)
in 1993 by Lakhtakia and Weiglhofer [1], the basis of realizing
these materials using thin--film technology was enunciated by Lakhtakia
and Messier [2]. Verification  by Robbie {\em et al.\/} [3]
soon followed, although a few years later I came across a pioneering and essentially comprehensive but obscure 
precedent reported by Young and Kowal  in 1959 [4]. The general concept
of sculptured thin films (STFs) emerged naturally, and was
presented in August 1995 by Lakhtakia and Messier to a 
group of thin--film researchers assembled at Penn State [5].
The topic has enjoyed considerable growth in the ensuing years,
chiefly in theory initially, but now in experiments and applications as well [6].

\skipline

The nanostructure of STFs comprises clusters of 3--5~nm
diameter and arranged
to form parallel columns that are bent in some fanciful forms with feature
size 30 nm or larger. Accordingly, a STF is a unidirectionally nonhomogeneous
continuum with direction--dependent properties at visible and infrared wavelengths.
A multi--section STF can thus be conceived of as an optical circuit
that can be integrated with electronic circuitry on a chip. Being porous,
a STF can act as a sensor of fluids and can be impregnated with
liquid crystals for switching applications too. Application as low--permittivity
barrier layers in electronic chips  as well
as for solar cells has also been suggested. During the last five
year, several physical vapor deposition techniques have emerged
for manufacturing STFs, and the first optical applications
saw the light of the day in 1999. 

\skipline

The following is a brief review of (i) the electromagnetic field equations, (ii) a nominal
nano\-scop\-ic--to--continuum model, (iii) realized optical applications,
and (iv) emerging applications of STFs. A large part of the work reviewed here
is due to my collaborators, my students and me. 
For the essentials of the
fabrication
techniques, the interested reader is enjoined to read the gem
of a paper that Young \& Kowal wrote [4]. For details of
the modern versions of the Young--Kowal technique, reference is
made
to Lakhtakia \& Messier [7],
Hodgkinson \& Wu [8],
Messier {\em et al.} [9], and Malac \&
Egerton [10]. The materials that can be deposited as STFs range from insulators
to semiconductors
to metals, thereby indicating the versatility of STF technology.

\skipline\skipline
{\bf 2. Electromagnetic field equations}

\skipline

{\it 2.1 Linear constitutive equations}

\skipline

Let the
$z$ axis of a cartesian coordinate system be aligned parallel to the direction
of non\-homo\-geneity. By definition,  the morphology of a 
simple STF in any plane $z = z_1$
can be made to coincide with the morphology in another plane $z=z_2$
with the help of a suitable rotation.  In other words, the {\em local\/}
morphology is spatially uniform, but the {\em global\/} morphology
is unidirectionally nonhomogeneous. Naturally, this leads to the concept
of local or {\em reference constitutive properties\/} of the STF. 
The global constitutive properties of the STF can be connected to the
local ones by means of rotation operators. 

\skipline

The frequency--domain
constitutive relations of a chosen STF are therefore defined as follows:
\begin{eqnarray}
\label{conreld1}
\DDr &=& \epso\,\=S(z)\. \Big[
\epsref\.\=S^{T}(z)\.\Er  + \alpharef\.\=S^{T}(z)\.\Hr\Big] \, ,\\
\label{conrelb1}
\Br &=& \muo\,\=S(z)\. \Big[
\betaref\.\=S^{T}(z)\.\Er  + \muref\.\=S^{T}(z)\.\Hr\Big] \, .
\end{eqnarray}
In these relations, $\epso = 8.854\times 10^{-12}$~F/m and
$\muo = 4\pi\times 10^{-7}$~H/m are the permittivity and the
permeability of vacuum, respectively, while the superscript $^T$ denotes
the transpose.
Whereas the 3$\times$3 dyadics $\epsref$ and $\muref$ represent the {\em reference\/}
dielectric and magnetic properties, respectively, the 3$\times$3 dyadics
$\alpharef$ and $\betaref$ delineate the {\em reference\/} magnetoelectric properties.
The
angular frequency is denoted by $\omega$, and an $\exp(-i\omega t)$
time--dependence is implicit.

\skipline

Any STF  for optical and/or infrared applications
comprises identical columns of 20--100~nm
diameter. Nominally, all columns twist
and bend identically as $z$ changes, which feature is captured
by the rotation dyadic $\=S(z)$. This 3$\times$3 dyadic
is some composition of the following
three elementary rotation 3$\times$3 dyadics:
\begin{eqnarray}
\=S_x(z) &=& \ux\ux + (\uy\uy + \uz\uz)\, \cos\, \xi(z) +
(\uz\uy - \uy\uz)\, \sin \,\xi(z)\, ,\\
\=S_y(z) &=& \uy\uy + (\ux\ux + \uz\uz)\, \cos\, \tau(z) +
(\uz\ux - \ux\uz)\, \sin \,\tau(z)\, ,\\
\=S_z(z) &=& \uz\uz + (\ux\ux + \uy\uy)\, \cos \,\zeta(z) +
(\uy\ux - \ux\uy)\, \sin \,\zeta(z)\, .
\end{eqnarray}
The angular functions of $z$ in these equations may be specified
piecewise, if necessary; while $\ux$, $\uy$ and $\uz$ are the
three cartesian unit vectors.

\skipline

All columns in a STF
are parallel to each other in any $xy$ plane, which characteristic is
incorporated in $\epsref$, $\muref$, $\alpharef$ and $\betaref$.
Suppose $\=S(0) = \=I$, the 3$\times$3 identity dyadic, so that
the plane $z=0$ is the reference plane. Let $\utau$ denote
a unit vector that is tangential at $z=0$ to any column.
Two other unit vectors are defined from the
shape of that column as follows:  the unit normal vector
$\un$ such that $\un\.\utau = 0$, and the unit binormal
vector $\ub =\utau\times\un$. 
Both $\utau$ and $\un$ are conveniently chosen to lie in the plane $y=0$; hence,
$\ub$ is either parallel or anti--parallel to $\uy$. With an angle
of rise $\chi$ specified, we  set
\begin{equation}
\utau = \ux\,\cos\,\chi + \uz\,\sin\,\chi\, ,\qquad
\un = -\,\ux\,\sin\,\chi + \uz\,\cos\,\chi\, ,\qquad
\ub =-\, \uy\, ,
\end{equation} 
these three unit vectors  forming
a right--handed coordinate system. Typically, the angle of rise $\chi \in (0^\circ,\, 90^\circ]$.

\skipline

Extensive experimental research on the uniform columnar thin films [11]
shows that the prescription
\begin{equation}
\label{biax}
\epsref = \epsa\, \un\un + \epsb\, \utau\utau + \epsc\,\ub\ub
\end{equation}
is appropriate. This $\epsref$ is a biaxial dyadic; and
the  simplification $\epsc = \epsa$
is appropriate for a locally uniaxial STF in the present
context.  If necessary, a gyrotropic term $\epsg\,\#u_g\times\=I$
may be appended to the right side of \r{biax}, with $\#u_g$ as some
unit vector. The forms of $\muref$, $\alpharef$ and $\betaref$ are similar
to those of $\epsref$. 
\skipline

Two canonical forms of STFs can be identified.  For sculptured nematic
thin films (SNTFs), either $\=S(z) = \=S_x(z)$ or $\=S(z) = \=S_y(z)$.
The columnar morphology is essentially 2--dimensional, lying
in either the $xz$ plane or the $yz$ plane. On the other hand,
thin--film helicoidal bianisotropic mediums (TFHBMs) are endowed
with 3--dimensional morphology, because $\=S(z) = \=S_z(z)$. Although TFHBMs
need not be periodically nonhomogeneous along the $z$ axis,
it is easy to fabricate them with periods chosen anywhere between
$200$~nm and $2000$~nm.
Of course,
combinations of the two canonical forms as well as cascades of multiple
sections are possible, and add to the attraction of
STFs.
\skipline

\skipline
{\it 2.2 Wave propagation}

\skipline

Electromagnetic wave propagation in a STF is best handled
using 4$\times$4 matrixes and column $4-$vectors. At any given
frequency,  the following spatial Fourier representation of the
electric and the magnetic field phasors is useful:
\begin{equation}
\left.\begin{array}{ll}
\#E(\#r,\omega) = \#e(z,\kappa,\psi_{inc},\omega) \,\exp\les i\kappa (x\cos\psi_{inc} + 
y\sin\psi_{inc})\ris\\ [6pt]
\#H(\#r,\omega) = \#h(z,\kappa,\psi_{inc},\omega)  \,\exp\les i\kappa (x\cos\psi_{inc} + 
y\sin\psi_{inc})\ris
\end{array}\ric\,.
\end{equation}
Substitution of the foregoing representation into the source--free Maxwell curl 
postulates, 
$\nabla\times\#E(\#r,\omega)
=i\omega\#B(\#r,\omega)$ and 
$\nabla\times\#H(\#r,\omega)
=-i\omega\#D(\#r,\omega)$, followed by the use of the constitutive relations
\r{conreld1} and \r{conrelb1} leads to four ordinary differential
equations and two algebraic equations. The components $e_z(z,\kappa,\psi_{inc},\omega)$
and $h_z(z,\kappa,\psi_{inc},\omega)$ are then eliminated to obtain the 4$\times$4 matrix
differential equation [12]
\begin{equation}
\label{wp}
\frac{d}{dz}\, [\#f(z,\kappa,\psi_{inc},\omega)] = i [\#P(z,\kappa,\psi_{inc},\omega)]
\, [\#f(z,\kappa,\psi_{inc},\omega)]\,.
\end{equation}
In this equation,
\begin{equation}
[\#f(z,\kappa,\psi_{inc},\omega)] = [e_x(z,\kappa,\psi_{inc},\omega),\,
e_y(z,\kappa,\psi_{inc},\omega),\,h_x(z,\kappa,\psi_{inc},\omega),
\,h_y(z,\kappa,\psi_{inc},\omega)]^T
\end{equation}
is a column $4-$vector, and $[\#P(z,\kappa,\psi_{inc},\omega)]$ is a 4$\times$4
matrix function of $z$ that can be easily obtained using symbolic manipulation
programs such as Mathematica.

\skipline

Analytic solution of \r{wp} can be obtained, provided $[\#P(z,\kappa,\psi_{inc},\omega)] =
[\#P_{con}(\kappa,\psi_{inc},\omega)]$ is not a function of $z$. This happens, of course,
for columnar thin films [11], and the solution procedure is described by Hochstadt [12].
Exact analytic solution of \r{wp} has been obtained also for axial
propagation (i.e., $\kappa=0$)
in periodic TFHBMs [14]; and a solution in terms of a convergent matrix polynomial series
is available for oblique propagation  (i.e., $\kappa \neq 0$) in periodic TFHBMs [15].

\skipline

More generally, only a numerical solution of \r{wp} can be obtained. Suppose 
that $[\#P(z,\kappa,\psi_{inc},\omega)]$ is a periodic function of $z$. Then, a perturbative 
approach can be used to obtain simple results for weakly periodic STFs [16, 17], while a 
coupled--mode approach may come handy if otherwise [18]. But if  
$[\#P(z,\kappa,\psi_{inc},\omega)]$ is not periodic, 
the constitutive dyadics can assumed to 
be 
piecewise homogeneous over slices of thickness $\Delta z$, and the 
approximate transfer equation [12]
\begin{equation}
	[\#f(z+\Delta z,\kappa,\psi_{inc},\omega)] \simeq \exp \lec i[\#P(z+\frac{\Delta 
z}{2},\kappa,\psi_{inc},\omega)] \,\Delta z\ric\,\,[\#f(z,\kappa,\psi_{inc},\omega)]
\end{equation}
can be suitably manipulated with appropriately small values of $\Delta z$.

\skipline\skipline
{\bf 3. From the nanoscopic to the continuum}

\skipline

{\it 3.1 Nominal model}

\skipline

Equations \r{conreld1} and \r{conrelb1}  incorporate the assumption of
a STF as a unidirectionally nonhomogeneous
continuum. This is valid in a macroscopic
sense, i.e., when the length scale of the film
morphology is considerably smaller than the electromagnetic
probe wavelength.   This assumption holds true
in the visible and the infrared frequency regimes,  because
STFs with appropriate morphological
length scales can be fabricated.  

\skipline

Let us consider a simple nanoscopic--to--macroscopic
homogenization formalism to determine the reference constitutive dyadics
$\epsref$, etc. [19]. The chosen STF is made of a bianisotropic material,
whose bulk constitutive relations are specified as
\begin{equation}
\left.\begin{array}{l}
\DDr =\epso\,  [
\=\eps_{\,s}(\omega)\.\Er  +\=\alpha_{\,s}(\omega)\.\Hr ] \\[6pt]
\Br = \muo\, [
\=\beta_{\,s}(\omega)\.\Er  + \=\mu_{\,s}(\omega)\.\Hr ] 
\end{array}\ric
\, .
\end{equation}
The void regions of the STF are taken to be occupied by a medium with the
following bulk properties:
\begin{equation}
\left.\begin{array}{l}
\DDr =\epso\,  [
\=\eps_{\,v}(\omega)\.\Er  +\=\alpha_{\,v}(\omega)\.\Hr ] \\[6pt]
\Br = \muo\, [
\=\beta_{\,v}(\omega)\.\Er  + \=\mu_{\,v}(\omega)\.\Hr ] 
\end{array}\ric
\, .
\end{equation}
Both mediums are supposed to be present in the STF as ellipsoids which
are entirely notional. In every $xy$ plane, the longest axes of all ellipsoids of
both mediums should be aligned in parallel. The shapes of the two types of ellipsoids
can be different, the respective surfaces being defined by
the functions
\begin{equation}
\#r_{s} (\theta, \phi) =  \delta_s \,\=U_{\,s}\.\ur (\theta,\phi)\, , \qquad
\#r_{v} (\theta, \phi) =  \delta_v \,\=U_{\,v}\.\ur (\theta,\phi)\, .
\end{equation}
Here, $\ur(\theta,\phi)$
is the radial unit vector in a spherical coordinate system located at the
centroid of an ellipsoid; the scalars $\delta_{s,v}$ are linear measures of the ellipsoidal
sizes; and the shape dyadics $\=U_{\,s,v}$  are real dyadics of rank 3, with  positive
eigenvalues  $0 < a_{s,v}^{(j)} \leq 1$, ($j = 1, \, 2, \, 3$). Setting $a^{(3)} \gg a^{(1)}$
and $a^{(3)} \gg a^{(2)}$ will make a particular ellipsoid almost like a needle with
a slight bulge in its middle part. The porosity of the
STF is denoted by $f_v$, ($0 \leq f_v \leq 1$). 

\skipline

The use of   6$\times$6 dyadics provides notational simplicity
for treating electromagnetic fields in bianisotropic materials. Thus, we define
the 6$\times$6 dyadics
\begin{equation}
 \*C_{\,ref,s,v} = \le \begin{array}{l|l} \epso\,\=\eps_{\,ref,s,v} & \epso\,\=\alpha_{\,ref,s,v} 
\\ 
\\ \hline \\
                 \muo\,\=\beta_{\,ref,s,v}      &  \muo\,\=\mu_{\,ref,s,v}  \end{array} \ri \, .
\end{equation}
The $\omega$--dependences of various
quantities are not explicitly mentioned in this
and the following equations
for compactness.

\skipline

The celebrated Bruggeman formalism is now implemented to 
effect local homogenization (with reference to any $xy$ plane)
[19].
For this purpose, the 6$\times$6 polarizability dyadics
\begin{equation}
\*a_{\,s,v} =  \le \*C_{\,s,v} - \*C_{\,ref} \ri \. \les \, \*I +
i\omega \,\*D_{\,s,v} \.
\le  \*C_{\,s,v} - \*C_{\,ref}  \ri \ris^{-1}
\end{equation}
are defined,
where $\*I$ is the 6$\times$6 identity dyadic. The 6$\times$6 depolarization
dyadics $\*D_{\,s,v}$ must be computed {\em via\/} two--dimensional
integration as follows:
\begin{eqnarray}
\nonumber
\*D_{\,s,v} &=& \frac{1}{4\pi i \omega\epso\muo}\, \int_{\phi=0}^{2\pi}\,d\phi\, 
\int_{\theta=0}^{\pi}\,d\theta\,  \sin\theta\, \frac{(\=U_{s,v}^{-
1}\.\ur)(\ur\.\=U_{s,v}^{\dag})}{A_{s,v}(\ur)}\\ \nonumber & & \\
\nonumber & & \\
& &\, \times\,
\le\begin{array}{l|l} \muo \ur\.\=U_{s,v}^{\dag}\.\=\mu_{\,ref}\.\=U_{s,v}^{-1}\.\ur & 
 -\,\epso\ur\.\=U_{s,v}^{\dag}\.\=\alpha_{\,ref}\.\=U_{s,v}^{-1}\.\ur \\ \\ \hline \\
-\,\muo \ur\.\=U_{s,v}^{\dag}\.\=\beta_{\,ref}\.\=U_{s,v}^{-1}\.\ur & 
 \epso\ur\.\=U_{s,v}^{\dag}\.\=\eps_{\,ref}\.\=U_{s,v}^{-1}\.\ur \end{array}\ri \, .
\end{eqnarray}
Here, $\ur=\ur(\theta,\phi)$ is the unit radial vector,
\begin{eqnarray}
\nonumber 
A_{s,v}(\ur)  &=&
(\ur\.\=U_{s,v}^{\dag}\.\=\eps_{\,ref}\.\=U_{s,v}^{-1}\.\ur )
(\ur\.\=U_{s,v}^{\dag}\.\=\mu_{\,ref}\.\=U_{s,v}^{-1}\.\ur ) \\
& &\qquad -\, (\ur\.\=U_{s,v}^{\dag}\.\=\alpha_{\,ref}\.\=U_{s,v}^{-1}\.\ur )
(\ur\.\=U_{s,v}^{\dag}\.\=\beta_{\,ref}\.\=U_{s,v}^{-1}\.\ur )
\end{eqnarray}
while $\=U_s^{\dag}$ is the transpose of $\=U_s^{-1}$, etc.
The Bruggeman formalism then consists of the solution of the
equation [20, 21]
\begin{equation}
\label{brug1}
f_v\,\*a_v + (1-f_v)\,\*a_s = \*0 \, ,
\end{equation}
with $\*0$ as the 6$\times$6 null dyadic. This equation
has to be numerically solved for $\*C_{\,ref}$, and a Jacobi iteration
technique is recommended for that purpose [21, 22]. 

\skipline

The solution of \r{brug1}  represents the homogenization of objects
of microscopic linear dimensions to a macroscopic continuum. The
quantities entering $\=S(z)$ are known prior
to fabrication, as also are $\chi$, $ \*C_{\,s}$
and  $\*C_{\,v}$. In order to calibrate the nominal model presented,
the shape dyadics $\=U_s$ and $\=U_v$ can
be chosen by comparison of the  predicted  $ \*C_{\,ref}$ against measured data.

\skipline
{\it 3.2 Application to dielectric TFHBMs}

\skipline

Sherwin and Lakhtakia [22] used the foregoing model to extensively study
the planewave responses of dielectric TFHBMs,
with $\=\mu_{\,s,v} =\=I$, $\=\alpha_{\,s,v} = \=0$,
$\=\beta_{\,s,v} = \=0$, $\=\eps_{\,s} = \eps_s\,\=I$, and
$\=\eps_{\,v} = \=I$. With $\lambdao$ denoting the 
wavelength in vacuum, a Lorentz
resonance model was chosen for $\eps_s$ as per
\begin{equation}
\epsilon_{s} = 1 +
\frac{q_{s}}{1+\left(N_{s}^{-1} - i \lambda_{s} \lambdao^{-1}
\right)^{2}}\, ,
\end{equation}
with constants $q_s$, $N_s$ and $\lambda_s$
selected
so that absorption is moderate for visible
wavelengths. The ellipsoids were chosen
to be identical (i.e., $\=U_s =\=U_v$), as described by
\begin{equation}
\label{E:one}
    (\#r\.\un)^2 + \left( \frac{\#r\.\ub}{\gamma_{2}} \right)^{2} + \left(
\frac{\#r\.\utau}{\gamma_{3}} \right)^{2} = \delta^{2}\,,
\end{equation}
where  the
transverse aspect ratio $\gamma_{2}>1$ and the slenderness ratio
$\gamma_{3} \gg1$ relate the three principal axes. 

\skipline

For normally incident planewaves and with
\begin{equation}
\label{ff}
\=S(z) =\=S_z(z)\Big\vert_{\zeta(z)=\pi z/\Omega}=
\uz\uz + (\ux\ux + \uy\uy)\, \cos \frac{\pi z}{\Omega} +
(\uy\ux - \ux\uy)\, \sin \frac{\pi z}{\Omega}\, ,
\end{equation}
the spectrums of optical
rotation, transmittance ellipticity,
linear dichroism,
circular dichroism, apparent linear dichroism,
and apparent circular
dichroism  were calculated as functions of the 
constitutive and
the geometric parameters $\eps_{s}$, $\gamma_{2}$, $\gamma_3$,
$\chi$, $\Omega$, $f=1-f_v$, and  $\lambdao$.
Maximum magnitudes in the
computed spectrums were determined in specific wavelength--regimes.
The variations of these maximums were then examined
with respect to any one of the constitutive and the geometric
parameters, while the other parameters were held fixed. 
From these studies, the following significant
conclusions were reached:
\begin{itemize}
\item[(i)] All observable response properties strongly depend on the
transverse aspect ratio $\gamma_{2}$, $1 \leq \gamma_{2} \ll
\gamma_{3}$. There exists a specific value of $\gamma_{2}$ denoted
by $\gamma_{2}^{o}$ such that all optical activity disappears. The
value of $\gamma_{2}^{o}$ can be parameterized in terms of other
geometric factors and $\eps_{s}$.

\item[(ii)] All observable property maximums are best--fitted to
fourth--order polynomials of $\gamma_{2}$.

\item[(iii)] All observable property maximums have similar
dependencies on the volume fraction $f$; furthermore, 
 $p^{max} \to 0$ as $f \to 0,1$. (Here, $p^{max}$
denotes the maximum magnitude  of the observable response
property $p$ over a prescribed range
of $\lambdao$.)

\item[(iv)] There exists an $f_{o}$ for each value of $\gamma_{2}$
such that $p^{max}(f_{o}) > p^{max} (f), f \ne f_{o}$. The value
of $f_{o}$ depends on the values of other geometric and
constitutive parameters as well as on the property $p$. An increase in
$\gamma_{2}$ results in a decrease in $f_{o}$.
\end{itemize}
Details of these and other  results will appear in print shortly [23], and the identified
 functional
relationships should assist in design of STF--based devices as
well as the on--line monitoring of STF fabrication processes.

\skipline\skipline
{\bf 4. Optical applications}

\skipline

Although many applications are
possible [5], the potential of STFs has been most successfully exploited
for optical filters. Chiral STFs, which are appropriately described as the periodic
dielectric TFHBMs of Section 3.2, must
display the circular Bragg phenomenon in accordance with
their periodic nonhomogeneity along the $z$ axis [6, 24], thereby ensconcing
themselves firmly in the area of {\em chiral optics\/}. Briefly, 
a structurally right-- (resp. left--) handed
chiral STF only a few periods thick almost completely reflects axially incident, right 
(resp. left) circularly polarized 
light with wavelength lying in the so--called Bragg regime; while the reflection
of axially incident, left (resp. right) circularly polarized light
in the same regime is very little.
The bandwidth of the Bragg regime and the peak reflectivity
therein first increase with the thickness of the chiral STF, and
then saturate. Once this saturation has occurred, 
further 
thickening of the film has negligible effects on the reflection spectrum.

\skipline
{\it 4.1 Circular polarization filters}

\skipline

The circular Bragg phenomenon  can be employed
to realize circular polarization filters. Normally incident, circularly polarized light of
one handedness can be reflected almost completely, while that of the other
handedness is substantially transmitted, if absorption is small enough and
the film is sufficiently thick,
in the Bragg regime. This was demonstrated by Wu {\em et al.} [25]
with chiral STFs fabricated using the serial bideposition technique. As of
now, the
Bragg regime can be positioned at virtually any $\lambdao \in [450,\, 1700]$~nm.
Polarization insensitivity can be realized using a bilayer version, as a cascade of
two otherwise identical chiral STFs but of opposite structural handedness
[26]. Chirping can be used to widen the bandwidth [27], and
tightly interlaced chiral STFs may also hold technological
attraction [28].

\skipline
{\it 4.2 Polarization--discriminatory handedness--inverters}

\skipline

A polarization--discriminatory handedness--inverter
for circularly polarized light  was fabricated using STF
technology. This was the first realization of a two--section STF device. It
comprises a chiral STF [24] and a half--wave plate realized as a
columnar thin film [11]. Basically, it almost completely reflects,
say, left circularly polarized light; while it
substantially transmits incident right circularly polarized
light after transforming it into left circularly polarized light. Theoretical
predictions [29] were borne out experimentally [30].

\skipline
{\it 4.3 Spectral hole filters}

\skipline

In a further bid to illustrate the potential of the STF concept,
a three--section STF was proposed as a spectral hole filter.
Its first and third layers are identical 
chiral STFs, while
the thin middle layer is homogeneous [31, 32].  The middle
layer is supposed to act as a phase defect. This design
was actually implemented to obtain a $11$~nm wide spectral
hole centered at $\lambdao = 580$~nm [33]. The realized bandwidth  
filter compares very favorably with those of commercially available
holographic filters.

\skipline

A better design became available shortly thereafter and was
experimentally evaluated too [34].
The middle layer was eliminated, but the lower chiral STF was
twisted by $90^\circ$ with respect to the upper chiral STF
about the $z$ axis. The twist acts as the required phase defect.

\skipline

{\it 4.4 Rugate and \u{S}olc filters}\\

SNTFs can also be pressed into service as optical filters~---~for linearly
polarized plane waves. McPhun {\em et al.\/} [35] fabricated rugate filters
with STF technology for narrow--band reflection applications. \u{S}olc
filters of the {\em fan} and the {\em folded} types are also possible with
the same technology [36].

\skipline

{\it 4.5 Fluid sensors}

\skipline

The porosity of STFs makes them attractive
for fluid concentration sensing applications [37, 38],
because their optical response properties must change in accordance with
the number density of molecules intruding into the void regions. In particular,
theoretical research has shown that
the Bragg regime of a chiral STF must shift accordingly,
thereby providing a measure of the fluid concentration [37]. Qualitative
support for this finding is provided by experiments on wet and dry chiral STFs
[39].

\skipline

Very recent theoretical research has indicated that STF spectral hole filters
can function as highly sensitive fluid concentration sensors; and
proof--of--concept experiments with both circularly polarized and
unpolarized incident light have confirmed the red--shift of spectral
holes upon exposure to moisture [40].

\skipline\skipline
{\bf 5. Emerging applications}

\skipline

From their inception [5], STFs were expected to have a wide range of applications, 
implementable only after their optical, electronic, and magnetic properties came to better 
understood. Their optical applications came to be investigated first, as detailed in Section 4. 
However, their high porosity~---~in combination with optical anisotropy and possible 
two--dimensional electron confinement in the microstructure~---~makes STFs potential candidates as 
\begin{itemize}
	
\item[(i)]	electroluminescent devices (emitting light of a pre--specified polarization state 
from circular to linear) prepared by chemical vapor deposition of nanocrystal silicon
into the void spaces of STFs formed from wide--gap transparent oxides; 
\item[(ii)]	high speed, high efficiency electrochromic films;
\item[(iii)]	optically transparent conducting films sculptured from pure metals; 
\item[(iv)]	multi--state electronic switches based on filamentary conduction; 
\item[(v)]	optical sensors that can detect and quantify various chemical and biological fluids; and
\item[(vi)]	micro--sieves for the entrapment of viruses or for growing biological tissues on 
surfaces of biological or non--biological provenances.
\end{itemize}
Obviously, many other applications may turn out to be possible, but significant progress thus far
has been reported, to my knowledge, only in the following five areas:

\skipline
{\it 5.1 Biochips}

\skipline

Endowed with porosity of engineered texture,  STFs can function as 
microreactors and therefore
can function as biochips. As an example, let us consider the following
scenario: 
Intercalation of a
ruthenium complex with double--stranded DNA
is known to generate luminescence. Suppose
that identical single--stranded DNA molecules~---~with a particular genomic
sequence matched to, say, {\em E. coli\/}~---~are
dispersed in a STF. A drop of contaminated
water containing analyte DNA molecules from exploded {\em E. coli\/} 
is put on the STF, followed by a drop of an appropriate
ruthenium complex. The bioluminescence signal emerging from
the STF can be optically sensed by photon counters in order to measure the degree
of contamination. 

\skipline

Bioluminescent emission is bound to be affected by
the reactor characteristics. If the reactor is a chiral STF,  the possibility of
exploiting the circular Bragg phenomenon exhibited by it  would be
attractive. Indeed, the structural handedness as well as
the periodicity of chiral STFs have been shown to critically control the emission
spectrum and intensity, while the polarization state of the emitted light is strongly correlated with the structural handedness of the embedded source filaments [41]. Bioluminescence STF
sensors therefore merit closer attention.

\skipline
{\it 5.2 Optical interconnects}

\skipline

Efficient use of optoelectronic devices requires
the development of optical interconnects which, in addition
to providing effective signal transmission, must be simple
to fabricate on integrated circuitry.  STF technology is compatible
with the planar technology of electronic chips.  Guided wave propagation
in chiral STFs turns out to yield the space--guide concept:
the capacity 
to simultaneously support propagation modes with different
phase velocities  in different
directions [42, 43].  This could 
result in efficient
use of the available {\it real estate\/} in
electronic chips. Furthermore, the helicoidal microstructure of chiral
STFs would resist
vertical cleavage and fracture, in contrast to columnar thin films which can also
function as space--guides.

\skipline

Chiral STFs can be grown as a regular lattice by lithographically
patterning the substrates [10]. Whereas slow substrate--rotation rates result
in the growth of arrays of microhelixes or microsprings spaced as close as 20~nm from their
nearest neighbors, faster rotation rates yield arrays of increasingly denser pillars
[9, 10]. Such STFs are essentially photonic bandgap materials
in the visible and the infrared regimes. Most recently, even line defects
have been introduced therein [44].

\skipline
{\it 5.3 Interlayer dielectrics}
\skipline
With the microelectronics industry moving relentlessly towards decreasing feature
sizes and increasingly stringent tolerance levels, an urgent need exists
for the use of low--permittivity materials as interlevel dielectrics. Silicon dioxide,
the current material of choice, has excellent properties in all respects except one:
its permittivity is too high. The porosity of STFs and nanoporous
silica makes them attractive
low--permittivity materials for microelectronic and electronic packaging
applications. However, chiral STFs are likely to have significant
thermal, mechanical, as well as electrical advantages
over nanoporous silica~---~because of (i) porosity with controllable
texture and (ii) helicoidal
morphology.
Also, chiral STFs could be impregnated with various kinds of polymers [45].

\skipline
{\it 5.4 Electrically addressable displays}
\skipline

Liquid crystals (LCs) can be electronically addressed [46] and are therefore widely
used these days for displays. Although STFs are not electronically addressable,
the alignment of nematic LCs forced into the void
regions of chiral STFs has been shown
to respond to applied voltages [47]. Thus, STF--LC
composites may have a future as robust displays.

\skipline
{\it 5.5 Optical pulse--shapers}
\skipline

The current explosive growth of digital optics 
communication has provided impetus for time--domain research on
novel materials. As chiral STFs are very attractive
for optical applications, the circular Bragg phenomenon is being
studied in the time domain. A pulse bleeding phenomenon has been
identified as the underlying mechanism, which can drastically affect
the shapes, amplitudes and spectral components of femtosecond pulses
[48]. However, narrow--band rectangular pulses can pass through
without significant loss of information [49].  Application of STFs
to shape optical pulses appears to be waiting in the wings.

\skipline\skipline
{\bf 6. Concluding remarks}

\skipline

The foregoing section makes it evident that
sculptured thin films are comparable to liquid crystals [45, 50, 51] in many
respects. Their respective roles can be competitive as well as
complementary, depending on the specific environment and
application. For instance, being soft
and viscous, LCs serve as pressure/temperature sensors.  But STFs are porous solids
and likely to be unaffected by small changes in the ambient pressure and temperature.
However, those qualities are likely to be useful in certain environments where
mechanical integrity and thermal stability are at a premium. Furthermore,
STFs can serve as fluid sensors and microreactors, but LCs can not.
Liquid crystals are widely used for electronic displays as they are electrically
addressable; STF--LC composites can also be electrically
addressed. Both LCs and STFs can be used as optical filters and polarizers,
but only the latter truly make the concept of optics--in--a--chip possible. Finally,
the wide scope of STFs~---~accessible through
tailorable microstructure and anisotropy, as well as {\em via} the
almost unlimited  types of depositable materials~---~is remarkable.

\skipline

To conclude, the development of STF technology is now in
a post--embryonic stage. Much needs to be done to make it
robust, economical and widely used. But the future appears bright,
and the recent feat of  Suzuki \& Taga [52] in fabricating STFs with many sections
underscores the tremendous promise of STFs as integrated optical chips [5, 12].

\skipline
\skipline
{\small {\bf Acknowledgements.} I am indebted to all of my collaborators
and my students for splendid support on STF research
over the last decade. This review
is dedicated to the inspirational
 batting performances of V.V.S. Laxman
and Rahul S. Dravid and the superb bowling performance of
Harbhajan Singh during a cricket Test match between India and Australia,
played March 11--15, 2001 at Eden Gardens, Kolkata.}

\skipline\skipline
{\bf 7. References}\\

[1]
A. Lakhtakia, W.S. Weiglhofer,
Microw. Opt. Technol. Lett.  6 (1993) 804.\\

[2]
A. Lakhtakia, R. Messier,
in: F. Mariotte, J.--P. Parneix (Eds.),
Proceedings of Chiral '94
 (P\'erigueux, France, May 18--20, 1994), French Atomic Energy Commission, Le Barp, 
France, 1994, pp. 125--130.\\

[3]
K. Robbie, M.J. Brett, A. Lakhtakia,
J. Vac. Sci. Technol. A 13 (1995) 2991.\\

[4]
N. O. Young, J. Kowal,
Nature  183 (1959) 104.\\

[5]
A. Lakhtakia, R. Messier, M.J. Brett, K. Robbie,
Innov. Mater. Res. 1 (1996) 165.\\

[6] V.C. Venugopal, A. Lakhtakia,
in: O.N. Singh, A. Lakhtakia (Eds.),
Electromagnetic Fields in Unconventional Materials
and Structures,
Wiley, New York, 2000, pp. 151--216.\\

[7]
A. Lakhtakia, R.F. Messier (Eds.),
Engineered Nanostructural Films and Materials,
SPIE, Bellingham, WA, USA,1999.\\

[8]
I. Hodgkinson, Q.H. Wu,
Appl. Opt. 38 (1999) 3621.\\

[9]
R. Messier, V.C. Venugopal, P.D. Sunal,
J. Vac. Sci. Technol. A 18 (2000) 1538.\\

[10]
M. Malac, R.F. Egerton,
Nanotechnology 12 (2001) 11.\\

[11]
I.J. Hodgkinson, Q.--h. Wu,
Birefringent Thin Films and Polarizing Elements,
World Scienti\-fic, Singapore, 1997.\\

[12] 
A. Lakhtakia,
Optik 107 (1997) 57.\\

[13]
H. Hochstadt,
Differential Equations~---~A Modern Approach,
Dover Press, New York, 1975, chap. 2.\\

[14]
A. Lakhtakia, W.S. Weiglhofer,
Proc. R. Soc. Lond. A 448 (1995) 419; erratums: 454 (1998) 3275.\\

[15]
A. Lakhtakia, W.S. Weiglhofer,
Proc. R. Soc. Lond. A 453 (1997) 93; erratums: 454 (1998) 3275.\\

[16]
V.A. Yakubovich, V.M. Starzhinskii,
Linear Differential Equations with Periodic Coefficients,
Wiley, New York, 1975.\\

[17]
W.S. Weiglhofer, A. Lakhtakia,
Optik 102 (1996) 111.\\

[18]
K. Rokushima, J. Yamakita,
J. Opt. Soc. Am. A 4 (1987) 27.\\

[19]
A. Lakhtakia, P.D. Sunal, V.C. Venugopal, E. Ertekin,
Proc. SPIE 3790 (1999) 77.\\

[20]
W.S. Weiglhofer, A. Lakhtakia, B. Michel,
Microw. Opt. Technol. Lett. 15 (1997) 263; 
erratum: 22 (1999) 221.\\

[21] B. Michel,
in: O.N. Singh, A. Lakhtakia (Eds.),
Electromagnetic Fields in Unconventional Materials
and Structures,
Wiley, New York, 2000, pp. 39--82.\\

[22]
J.A. Sherwin, A. Lakhtakia,
Proc. SPIE 4097 (2000) 250.\\

[23]
J.A. Sherwin, A. Lakhtakia,
Math. Comput. Model. (accepted for publication in 2001).\\

[24]
I. Hodgkinson, Q.h. Wu, B. Knight, A. Lakhtakia, K. Robbie,
Appl. Opt. 39 (2000) 642.\\

[25]
Q. Wu, I.J. Hodgkinson, A. Lakhtakia,
Opt. Eng. 39 (2000) 1863.\\

[26]
A. Lakhtakia, V.C. Venugopal,
Microw. Opt. Technol. Lett. 17 (1998) 135.\\

[27]
A. Lakhtakia,
Microw. Opt. Technol. Lett. 28 (2001) 323.\\

[28]
A. Lakhtakia,
Optik 112 (2001) 119.\\

[29]
A. Lakhtakia,
Opt. Eng. 38 (1999) 1596.\\

[30]
I.J. Hodgkinson, A. Lakhtakia, Q.h. Wu,
Opt. Eng. 39 (2000) 2831.\\

[31]
A. Lakhtakia, M. McCall,
Opt. Commun. 168 (1999) 457.\\

[32]
A. Lakhtakia, V.C. Venugopal, M.W. McCall,
Opt. Commun. 177 (1999) 57.\\

[33]
I.J. Hodgkinson, Q.h. Wu, A. Lakhtakia, M.W. McCall,
Opt. Commun. 177 (2000) 79.\\

[34]
I.J. Hodgkinson, Q.H. Wu, K.E. Thorn, A. Lakhtakia, M.W. McCall,
Opt. Commun. 184 (2000) 57.\\

[35]
A.H. McPhun, Q.H. Wu, I.J. Hodgkinson,
Electron. Lett. 34 (1998) 360.\\

[36]
A. Lakhtakia,
Opt. Eng. 37 (1998) 1870.\\

[37]
A. Lakhtakia,
Sensors \& Actuators B: Chem. 52 (1998) 243.\\

[38]
E. Ertekin, A. Lakhtakia,
Eur. Phys. J. Appl. Phys. 5 (1999) 45.\\

[39]
I.J. Hodgkinson, Q.h. Wu, K.M. McGrath,
Proc. SPIE 3790 (1999) 184.\\

[40]
A. Lakhtakia, M.W. McCall, J.A. Sherwin, Q.h. Wu, I.J. Hodgkinson,
Opt. Commun. (accepted for publication in 2001).\\

[41]
A. Lakhtakia,
Opt. Commun. 188 (2001) 313.\\

[42]
A. Lakhtakia, 
Optik 110 (1999) 289.\\

[43]
E. Ertekin, A. Lakhtakia, 
Proc. R. Soc. Lond. A 457 (2001) 817.\\

[44]
M. Malac, R.F. Egerton,
J. Vac. Sci. Technol. A 19 (2000) 158.\\

[45]
V.C. Venugopal, A. Lakhtakia, R. Messier, J.--P. Kucera,
J. Vac. Sci. Technol. B 18 (2000) 32.\\

[46]
S.D. Jacobs (Ed.),
Selected Papers on Liquid Crystals for Optics,
SPIE, Bellingham, WA, USA, 1992.\\

[47]
J.C. Sit, D.J. Broer, M.J. Brett,
Liq. Cryst. 27 (2000) 387.\\

[48]
J.B. Geddes III, A. Lakhtakia,
Eur. Phys. J. Appl. Phys. 13 (2001) 3.\\

[49]
J.B. Geddes III, A. Lakhtakia,
Microw. Opt. Technol. Lett. 28 (2001) 59.\\

[50]
S. Chandrasekhar,
Liquid Crystals,
Cambridge University Press, Cambridge, UK, 1992.\\

[51]
P.G. de Gennes, J. Prost,
The Physics of Liquid Crystals,
Clarendon Press, Oxford, UK, 1993.\\

[52]
M. Suzuki, Y. Taga,
Jap. J. Appl. Phys. 40, part 2 (2001) L358.\\

\end{document}